# ACCOUNTING FOR SELF-PROTECTIVE RESPONSES IN RANDOMIZED RESPONSE DATA FROM A SOCIAL SECURITY SURVEY USING THE ZERO-INFLATED POISSON MODEL

By Maarten J. L. F. Cruyff, Ulf Böckenholt, Ardo van den Hout and Peter G. M. van der Heijden

*Utrecht University, McGill University and MRC Biostatistics Unit*

In 2004 the Dutch Department of Social Affairs conducted a survey to assess the extent of noncompliance with social security regulations. The survey was conducted among 870 recipients of social security benefits and included a series of sensitive questions about regulatory noncompliance. Due to the sensitive nature of the questions the randomized response design was used. Although randomized response protects the privacy of the respondent, it is unlikely that all respondents followed the design. In this paper we introduce a model that allows for respondents displaying self-protective response behavior by consistently giving the nonincriminating response, irrespective of the outcome of the randomizing device. The dependent variable denoting the total number of incriminating responses is assumed to be generated by the application of randomized response to a latent Poisson variable denoting the true number of rule violations. Since self-protective responses result in an excess of observed zeros in relation to the Poisson randomized response distribution, these are modeled as observed zero-inflation. The model includes predictors of the Poisson parameters, as well as predictors of the probability of self-protective response behavior.

**1. Introduction.** In 2004 the Dutch Department of Social Affairs conducted a nationwide survey to assess the level of compliance with the Unemployment Insurance Act. Under this act employees who have lost their income due to unemployment are entitled to financial benefits, provided that they comply with the rules and regulations stipulated in the act. The participants in the survey were asked if they had ever violated against the regulations in the year preceding the survey. Since the disclosure of a rule violation may have serious financial consequences for the respondent, the randomized response design was used.

The randomized response method was first introduced in 1965 by Warner as an interview technique that protects the respondents' privacy [Warner (1965)]. In Warner's design the respondent is presented with two complementary statements, for example "I am a marihuana user" and "I am not a marihuana user." The respondent then operates a randomizing device, like a pair of dice or a deck of cards, and the outcome of this device determines which of the two statements the respondents has to answer. Since only the respondent knows the outcome of the randomizing device, confidentiality is guaranteed.

A meta-analysis of randomized response studies shows that the randomized response design generally yields higher and more valid prevalence estimates of the sensitive characteristic than direct-questioning designs [Lensvelt-Mulders et al. (2005)]. However, a number of studies suggest that respondents do not always follow the instructions of the randomized response design. In an experimental randomized response design [Edgell, Himmelfarb and Duncan (1982)] with the outcomes of the randomizing device fixed in advance, about 25% of the respondents answers *no* to a question about having had homosexual experiences, while according to the design these respondents should have answered *yes*. In another experimental study [van der Heijden et al. (2000)] all respondents were known to have offended against social security regulations. Although the randomized response condition yielded higher estimates than the direct question design, the prevalence estimate of offenders obtained with randomized response was only about 50%. Another study involved an interview of participants in a randomized response survey [Boeije and Lensvelt-Mulders (2002)]. Many of the participants indicated that they had found it difficult to falsely incriminate themselves when they were forced to do so by the outcome of the dice. Some of them admitted that in this situation they had given the nonincriminating answer instead.

A recent topic of investigation in the field of randomized response is the estimation of evasive response bias. Clark and Desharnais (1998) show that the presence of evasive responses can be detected in a randomized response design with two groups that each use a randomizing device with different outcome probabilities. Kim and Warde (2005) present a multinomial randomized response model taking evasive response bias into account in designs with a sensitive question with multiple response categories that increase in sensitivity. The term self-protection (SP) was introduced by Böckenholt and van der Heijden (2004, 2007) to describe the responses by respondents who consistently give the evasive answer, without taking the outcome of the randomizing device into account. According to this definition, the SP response profile consists of nonincriminating (i.e., *no*) responses only. The authors use models from item response theory to obtain prevalence estimates of the sensitive characteristics corrected for SP. The SP assumption is also used in log-linear randomized response models that study the association patterns between the sensitive characteristics and obtain prevalence estimates corrected for SP [Cruyff, van den Hout, van der Heijden (2007)].

The definition of SP implies that the probability of an evasive response does not explicitly depend on the sensitivity of the question or on the true status of the respondent. Although it is possible to formulate more complex assumptions with respect to the generation of evasive response bias, SP seems to provide an adequate description of the process. A study [Böckenholt, Barlas and van der Heijden (2008)] modeling evasive response behavior in randomized response as a function of both the sensitivity of the question and the true status of the respondent found no compelling evidence for the superiority of these models in relation to the corresponding SP models.

In this paper we introduce a regression model that allows for SP in randomized response sum score data. The model assumes a Poisson distribution for the true sum score variable assessing the individual number of sensitive characteristics. The model further assumes that the observed sum score variable denoting the number of incriminating responses is partly generated by the randomized response design, and partly by SP. Since SP by definition results in an observed sum score of zero, the distribution of the observed sum score variable is zero-inflated with respect to the Poisson randomized response distribution of the true sum score variable. The model allows for predictors that explain individual differences in the Poisson parameters, as well as predictors that explain individual differences in the probability of SP. Since the distribution of the observed sum score variable is a mixture of a Poisson randomized response distribution and observed zero-inflation, the model is called the zero-inflated Poisson randomized response regression model.

The model is applied to randomized response data from a social security survey conducted in the Netherlands in 2004. Section 2 describes the data. Section 3 derives the zero-inflated Poisson regression model based as an extension of existing randomized response models for multinomial and sum score data. The section also includes a description of a maximum likelihood (ML) estimation procedure and an evaluation of the validity of the Poisson assumption with respect to the true sum score variable. The results for the social security data [Cruyff, Bckenholt, van den Hout and van der Heijden (2008b)] are presented in Section 4. Section 5 discusses some assumptions and interpretations of the model.

**2. The data.** In 2004 the Department of Social Affairs in the Netherlands conducted a nationwide survey to assess the level of noncompliance with the Social Security Law [compare Lensvelt-Mulders et al. (2006)]. The survey includes 870 participants who receive financial benefits under the Unemployment Insurance Act (UIA). Persons who have become (partially) unemployed are eligible for benefits. A beneficiary receives about 70% of the last earned wages, and the duration of the benefits depend on the length of the persons' employment history. Beneficiaries are required to report all activities that generate income in addition to their benefits or that might conflict with reintegration into the labor market. The failure to report such an activity may be sanctioned.

The social security survey includes the following five questions assessing noncompliance with UIA regulations:

1. Have you in the past 12 months ever had a job or worked for an employment agency in addition to your benefit without informing the Department of Social Services?
2. Have you in the past 12 months ever refused to accept a suitable job, or have you ever deliberately made sure you were not hired even though you had a chance of getting the job?
3. Have you in the past 12 months ever deliberately put in an insufficient number of job applications for a sustained period of time?
4. Have you in the past 12 months attended any day courses without informing the Department of Social Services?
5. Have you in the past 12 months had any income in addition to your benefit, for example, from alimony, a scholarship, subletting, other benefits, gifts, interest and so forth, without informing the Department of Social Services?

Due to the sensitive nature of the questions, the randomized response method is used. The respondents answer the questions with the use of a computer according to the forced response design [Boruch (1971)]. Before answering the question the respondent throws two virtual dice, and is instructed to answer *yes* if the sum of the dice is 2, 3 or 4, and to answer *no* if the sum of the dice is 11 or 12. If the sum of the dice is 5, 6, 7, 8, 9 or 10, the respondent has to answer the question truthfully. The misclassification probabilities, that are conditional on the true status of the respondent, can be derived from the probability distribution of the sum of two dice. Given regulatory noncompliance, the probability of a *yes* response is 11/12 and that of *no* response 1/12. Given regulatory compliance, the probability of a *yes* response is 1/6 and the probability of a *no* response 5/6. In the actual social security survey, however, the programmer inadvertently programmed the virtual dice so that the probability of a *yes* response given regulatory noncompliance was 0.9329, and that of a *yes* response given regulatory compliance 0.18678. The number of observed *yes* responses to the five questions are respectively 122, 195, 168, 207 and 274. Counting the total number of *yes* responses for each respondent on the five questions yields the frequencies $n_0 = 288$, $n_1 = 295$, $n_2 = 207$, $n_3 = 68$, $n_4 = 7$ and $n_5 = 5$ (with the subscript denoting the number of observed *yes* responses).

The social security survey includes two kinds of predictors we like to explore, one concerning demographic variables and the other concerning variables related to the forced response design. The demographic variables *gender*, *age*, *year unemployment*, *education* and *knowledge rules* are used as predictors of regulatory noncompliance. The variables *gender* and *age* are dummy-coded with "male" ($n = 483$) and "older than 26" ($n = 832$) as respective reference categories. The variable *year unemployment* is a dummy variable denoting the last year of being employed, with the year 2004 as reference category ($n = 257$). The variable *education* (mean $= 2.25$, sd $= 0.67$) measures increasing levels of eduction. The variable *knowledge rules* (mean $= 3.8$, sd $= 0.90$) denotes on a 5-point scale of the respondents' general knowledge of the social security regulations. The two variables *trust* and *understanding* are related to the forced response design and are used as predictors of SP. The variable *trust* (mean $= 3.5$, sd $= 0.92$) is constructed as the average score on four 5-point scale variables (Cronbach's Alpha $= 0.87$) assessing different aspects of the respondents' beliefs in the confidentiality and privacy protection of the forced response design. A high score on this variable corresponds to a high degree of trust. The variable *understanding* (mean $= 4.2$, sd $= 0.85$) assesses on a 5-point scale to what extent the respondent feels that he understood when to answer *yes* and when to answer *no* to a forced response question. High scores correspond to a good understanding of the forced response design.

Figure 1 depicts the associations between the observed sum scores and the predictors. At this point we would like to emphasize that the plots should not be interpreted as depicting associations between the predictors and the true sum scores (i.e., the number of rule violations), since the observed sum scores are not corrected for the misclassification due to randomized response, nor for SP. The plots at the top of the figure show the observed sum score proportion conditional on the categories within the dummy variables *gender*, *age* and *year unemployment*. The profiles of males and females look similar. The plot for *age* shows that the proportion of zeros for the younger respondents (about 15%) is about half that of the older respondents. The younger respondents also have a relatively high proportion of ones (about 45%). The profiles within *year unemployment* are again relatively similar, although persons who became unemployed in 2004 have a higher percentage of zero response (about 40%) compared to the respondents who became unemployed before 2004 (30%). The four plots at the bottom show the mean predictor scores within the observed sum scores for the respective continuous variables *education*, *knowledge rules*, *trust* and *understanding*. The predictor means do not show any clear linear associations with the observed sum score, although for the variable *knowledge rules* the means seem to slightly decrease with increasing sum scores. The effect of sum score is most pronounced on the means of the predictor *trust*, but the association pattern is erratic.

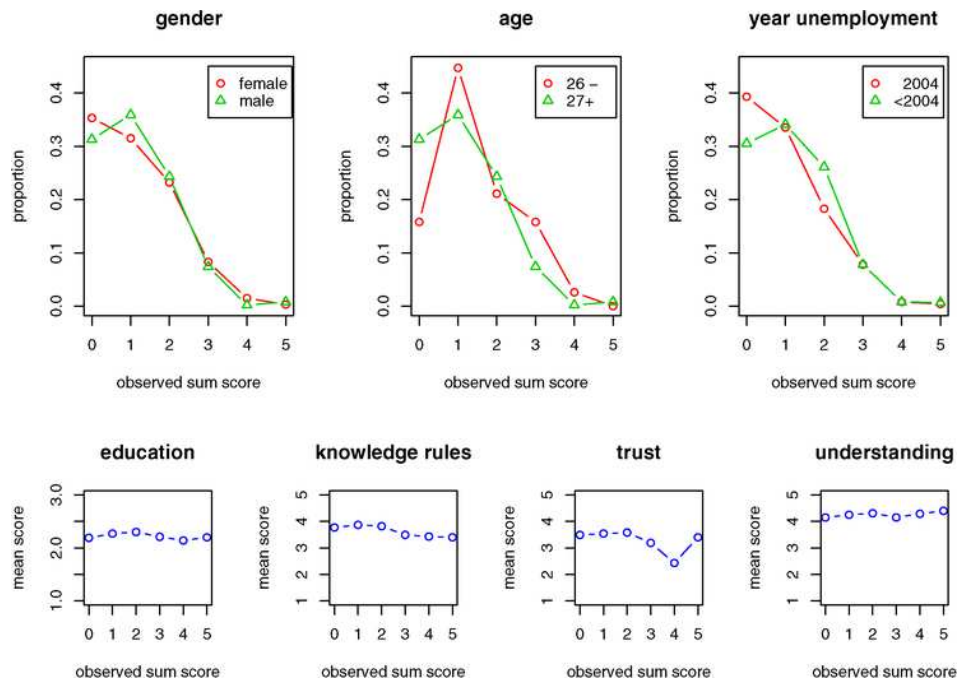

Fig. 1. *Observed sum score proportion given the categories within the dummy variables gender, age and year unemployment (upper plots), and mean predictor score within observed sum score for the continuous variables education, knowledge rules, trust and understanding (lower plots).*

## 3. The model.

3.1. *The multinomial randomized response model.* Consider a randomized response design with $M$ sensitive questions, each assessing the presence or absence of a sensitive characteristic. Let the random variable $Y_m^*$ denote the observed response to the $m$th question, with $y_m^* \in \{0 \equiv no, 1 \equiv yes\}$ and $m \in \{1, \ldots, M\}$. Similarly, let $Y_m$ denote the true status with respect to the sensitive characteristic, with $y_m \in \{0 \equiv \text{absent}, 1 \equiv \text{present}\}$. The binomial randomized response model for the binary variable $Y_m^*$ is given by

$$(1) \qquad \mathbb{P}(Y_m^* = y_m^*) = \sum_{y_m=0}^{1} p_{y_m^*|y_m} \pi_{y_m},$$

where $\pi_{y_m} = \mathbb{P}(Y_m = y_m)$ and $p_{y_m^*|y_m} = \mathbb{P}(Y_m^* = y_m^*|Y_m = y_m)$ are the conditional misclassification probabilities that can be derived from the probability distribution of the randomizing device. For a more detailed discussion of this model, we refer to Chaudhuri and Mukerjee (1988).

Next consider the true sum score of the $M$ sensitive characteristics, denoted by the variable

$$(2) \qquad S = \sum_{m=1}^{M} Y_m.$$

If $S$ follows a multinomial distribution with parameters $\pi_0, \ldots, \pi_M$, then the multinomial randomized response model

$$(3) \qquad \mathbb{P}(S^* = s^*) = \sum_{s=0}^{M} q_{s^*|s}\pi_s,$$

applies, where $S^*$ denotes the number of observed *yes* responses on the $M$ sensitive questions and $q_{s^*|s} = \mathbb{P}(S^* = s^*|S = s)$, for $s, s^* \in \{0, \ldots, M\}$.

The misclassification probabilities $q_{s^*|s}$, that exist if and only if the $p_{y_m^*|y_m}$ are the same for all $m$, can be derived as the multinomial probabilities

$$(4) \qquad q_{s|t} = \sum_{j=0}^{t} \binom{t}{j}\binom{M-t}{s+j-t} p_{1|1}^{t-j} p_{0|1}^{j} p_{1|0}^{s+j-t} p_{0|0}^{M-s-j},$$

for $s, t \in \{0, 1, \ldots, M\}$ and $t \le s + j \le M$ [Cruyff, van den Hout, van der Heijden (2008b)].

As an illustration, consider the forced response design of the social security survey with two binary variables $Y_1$ and $Y_2$. Application of (4) for $M = 2$ yields the misclassification probabilities $q_{s^*|s}$:

$$\begin{pmatrix} q_{0|0} & q_{0|1} & q_{0|2} \\ q_{1|0} & q_{1|1} & q_{1|2} \\ q_{2|0} & q_{2|1} & q_{2|2} \end{pmatrix} = \begin{pmatrix} p_{0|0}^2 & p_{0|0}p_{0|1} & p_{0|1}^2 \\ 2p_{0|0}p_{1|0} & p_{1|0}p_{0|1} + p_{0|0}p_{1|1} & 2p_{0|1}p_{1|1} \\ p_{1|0}^2 & p_{1|0}p_{1|1} & p_{1|1}^2 \end{pmatrix}.$$

3.2. *The Poisson randomized response model.* Assume that the true sum score $S$ is generated by a Poisson process with parameter $\lambda$. Since realizations of $S$ are limited to the maximum value of $M$, the Poisson distribution of $S$ is truncated at the right [Cameron and Trivedi (1998)], so that

$$(5) \qquad \mathbb{P}(S = s|s \le M) = \frac{\pi_s}{\sum_{s=0}^{M} \pi_s},$$

with

$$(6) \qquad \pi_s = \frac{\exp(-\lambda)\lambda^s}{s!}.$$

Substitution of the multinomial probabilities $\pi_s$ in model (3) for the expression at the right-hand side of (5), with $\pi_s$ defined as in (6), yields the (right-truncated) Poisson randomized response model

$$(7) \qquad \mathbb{P}(S^* = s^* | s^*, s \leq M) = \sum_{s=0}^{M} q_{s^*|s} \frac{\pi_s}{\sum_{s=0}^{M} \pi_s}.$$

3.3. *The zero-inflated randomized response regression model.* Count data are often characterized by an excess of zeros relative to a Poisson distribution. To account for the excess of zeros, Lambert (1992) introduced the zero-inflated Poisson (ZIP) model

$$(8) \qquad \mathbb{P}(S = s) = (1 - \theta)\pi_s + I\theta,$$

with $S \in \{0, 1, 2, \ldots\}$, $\pi_s$ defined as in (6), and $I$ an indicator variable taking on value 1 if $S = 0$, and 0 otherwise. The parameter $\theta$ denotes the probability of an excess zero in the observed counts, that is, a zero count that is not generated by the Poisson process.

Now suppose that in the context of randomized response the true sum score variable $S$ is generated by a Poisson process. In the absence of SP the observed sum score variable $S^*$ is entirely generated by the Poisson randomized response process. In the presence of SP, however, $S^*$ is generated partly by the Poisson randomized response process and partly by SP. Let the parameter $\theta^*$ denote the probability that the observed sum score is generated by SP, and let $1 - \theta^*$ denote the probability that the observed sum score is generated by a Poisson randomized response process. The distribution of $S^*$ is then given by

$$(9) \qquad \mathbb{P}(S^* = s^* | s^*, s \leq M) = (1 - \theta^*) \sum_{s=0}^{M} q_{s^*|s} \frac{\pi_s}{\sum_{s=0}^{M} \pi_s} + I^* \theta^*,$$

where $I^*$ is an indicator variable taking on the value 1 if $S^* = 0$, and 0 otherwise.

Both parameters $\lambda$ and $\theta^*$ in (9) can be modeled as a function of predictors. Let variable $S_i$ denote the true sum of sensitive characteristics of individual $i$, for $i \in \{1, \ldots, n\}$, and let $\mathbf{x}_i = (x_{i0}, \ldots, x_{ik})'$ and $\mathbf{z}_i = (z_{i0}, \ldots, z_{il})'$ be vectors that may or may not contain the same predictors. Let the Poisson parameter of individual $i$ depend on $\mathbf{x}_i$ according to

$$(10) \qquad \lambda_i = \exp(\mathbf{x}_i' \boldsymbol{\beta}),$$

and let the probability of zero-inflation depend on $\mathbf{z}_i$ according to

$$(11) \qquad \theta_i^* = \frac{\exp(\mathbf{z}_i' \boldsymbol{\gamma})}{1 + \exp(\mathbf{z}_i' \boldsymbol{\gamma})},$$

where $\boldsymbol{\beta} = (\beta_0, \ldots, \beta_k)'$ and $\boldsymbol{\gamma} = (\gamma_0, \ldots, \gamma_l)'$ are parameter vectors. The ZIP randomized response regression model is given by

$$(12) \qquad \mathbb{P}(S_i^* = s_i^* | s_i^*, s_i \leq M, \mathbf{x}_i, \mathbf{z}_i) = (1 - \theta_i^*) \sum_{s_i=0}^{M} q_{s_i^* | s_i} \frac{\pi_{s_i}}{\sum_{s_i=0}^{M} \pi_{s_i}} + I^* \theta_i^*,$$

where $\pi_{s_i} = \exp(-\lambda_i) \lambda_i^{s_i} / s_i!$.

3.4. *Estimation.* The ZIP randomized response regression model (12) (as well as the other models presented in this section) can be estimated by maximizing the kernel of the observed-data log-likelihood

$$(13) \qquad \begin{aligned} &\ln \ell^*(\boldsymbol{\beta}, \boldsymbol{\gamma} | \mathbf{s}^*, \mathbf{X}, \mathbf{Z}) \\ &= \sum_{i=1}^{n} \ln \left( (1 - \theta_i^*) \sum_{s_i=0}^{M} q_{s_i^* | s_i} \frac{\pi_{s_i}}{\sum_{s_i=0}^{M} \pi_{s_i}} + I^* \theta_i^* \right), \end{aligned}$$

with respect to the parameters $\boldsymbol{\beta}$ and $\boldsymbol{\gamma}$. We have written code for the quasi Newton–Raphson procedure `QNewtonmt` of the statistical software program GAUSS to estimate the model parameters. The procedure uses the BFGS method with numerically computed gradients and Hessian matrix, and standard errors are obtained from the inverse of the estimated Hessian. Convergence is generally fast, but due to machine imprecision problems may be encountered with the inversion of the Hessian. The use of slightly different starting values usually solves this problem. The observed-data likelihood is convex and unimodal when evaluated as a function of the parameters $\theta$ and $\lambda$. Figure 2 shows the shape likelihood function for the ZIP randomized response model 9 given the social security data. This model does not include any predictors for the parameters $\lambda$ and $\theta$, and the likelihood function is evaluated for the SP parameter $\theta \in (0, 0.25)$ and the Poisson parameter $\lambda \in (0.25, 0.75)$.

3.5. *The Poisson assumption.* It is a well known statistical result that for $M \gg 1$, $\pi \ll 1$ and $M\pi \approx 1$, the distribution of the sum of $M$ i.i.d. Bernoulli variables with success probability $\pi$ is approximated by a Poisson distribution with parameter $\lambda = M\pi$. The Poisson randomized response models presented in this paper are based on the assumption that the five randomized response variables $Y_m$ are Bernoulli variables and that the sum of these variables follows a Poisson distribution with parameter $\lambda = \sum_{m=1}^{M} \pi_{1_m}$, where $\pi_{1_m}$ denotes the success probability of variable $Y_m$ (i.e., the prevalence of the sensitive characteristic). In this section we evaluate the validity of this assumption, given that in our example $M$ is relatively small and that the success probabilities $\pi_{1_m}$ are not identical for different $m$.

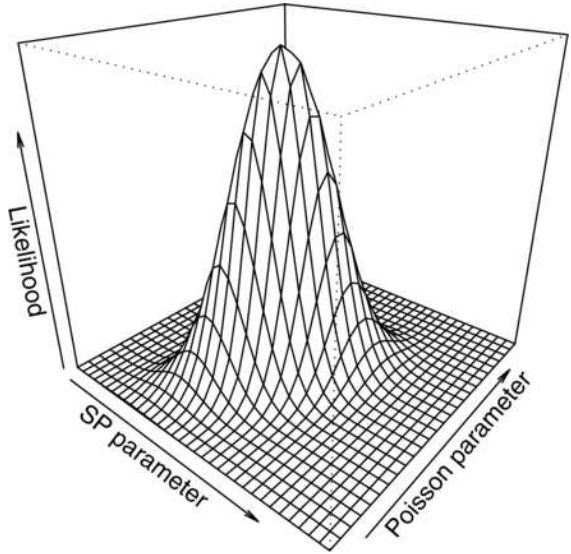

Fig. 2.    *Likelihood function for the ZIP randomized response model evaluated for the SP parameter $\theta^* \in (0, 0.25)$ and the Poisson parameter $\lambda \in (0.25, 0.75)$.*

TABLE 1

|  | 0 | 1 | 2 | 3 | 4 | 5 |
|---|---|---|---|---|---|---|
| Exact distribution | 0.7250 | 0.2498 | 0.0244 | 0.0008 | 0.0000 | 0.0000 |
| Poisson approximation | 0.7401 | 0.2228 | 0.0335 | 0.0033 | 0.0002 | 0.0000 |

To evaluate the adequacy of the Poisson assumption, we first derive the exact distribution of the sum of five independent Bernoulli variables with success probabilities equal to the prevalence estimates $\hat{\pi}_{1_m}$ of the five variables $Y_m$ of the social security survey. The prevalence estimates obtained with the multinomial randomized response model (1) are $\hat{\pi}_{1_1} < 0.001$, $\hat{\pi}_{1_2} = 0.050$, $\hat{\pi}_{1_3} = 0.009$, $\hat{\pi}_{1_4} = 0.069$ and $\hat{\pi}_{1_1} = 0.172$, with $\hat{\pi}_{1_1}$ set equal to 0.001. We then approximate this distribution by a Poisson distribution with $\lambda = \sum_{m=1}^{5} \hat{\pi}_{1_m} = 0.301$. The two distributions are shown in the Table 1.

The Poisson approximation assigns more mass to the zero count and to counts larger than 2, and thus overestimates the true variance. It underestimates the probability of count 1 by 0.0270, which corresponds to a relative difference of approximately 11%. In view of the fact that the absolute deviations in probability of the remaining counts are smaller, the Poisson approximation seems satisfactory for all practical purposes.

**4. Analysis of the social security data.** Table 2 presents fit indices for the multinomial randomized response model ($\mathcal{M}$), the Poisson randomized response model ($\mathcal{P}$), the ZIP randomized response null-model ($\mathcal{Z}_0$), the model $\mathcal{Z}_\beta$ including the five demographic predictors of the Poisson parameter, and the full model $\mathcal{Z}_{\gamma,\beta}$ with the additional two predictors of SP. The table reports the loglikelihood, the Akaike Information Criterion (AIC) given by $2k - 2\ln\ell^*$, the Bayesian Information Criterion (BIC) given by $k\ln n - 2\ln\ell^*$, with $\ln\ell^*$ the maximized loglikelihood and $k$ the number of independently estimated parameters. For the models without predictors, we present the Pearson chi-square statistic $X^2$ with $df = M - k$, where $M$ denotes the number of independently observed sum score frequencies. The last column of Table 2 presents the SP probability estimates $\widehat{\theta}^* = (\sum_{i=1}^n I_{(S_i^*=0)})^{-1} \sum_{i=1}^n I_{(S_i^*=0)} \widehat{\theta}_i$ for the three ZIP models.

Although model $\mathcal{M}$ is saturated, the fitted response frequencies $\widehat{n}_0 = 272.0$, $\widehat{n}_1 = 319.1$, $\widehat{n}_2 = 195.3$, $\widehat{n}_3 = 66.7$, $\widehat{n}_4 = 12.6$, $\widehat{n}_5 = 4.3$ do not equal the corresponding observed response frequencies. The fact that $X^2$ is nonzero with zero degrees of freedom indicates that one or more of the estimates are on the boundary of the parameter space [van den Hout and van der Heijden (2002)]. The expected distribution of the true sum score variable $S$ is

$$\widehat{\boldsymbol{\pi}}(\mathcal{M}) = (0.878, 0.000, 0.116, 0.000, 0.000, 0.006),$$

with (near) zero-probability estimates for one, three and four rule violations. An interesting result is that the probability estimate of 0.6% for five rule violations is inconsistent with the fact the smallest univariate prevalence estimate of regulatory noncompliance is only 0.1% (see Section 3.5).

Model $\mathcal{P}$ clearly does not fit well, indicating that for our application the Poisson assumption does not hold. SP is introduced in model $\mathcal{Z}_0$ with an estimated probability of 12.6%. This model fits substantially better and is the best model in terms of BIC. The Pearson chi-square of 19.6 with 4 degrees of freedom, however, indicates lack of fit. The fitted frequencies $\widehat{n}_0 = 287.2$, $\widehat{n}_1 = 298.9$, $\widehat{n}_2 = 199.5$, $\widehat{n}_3 = 70.1$, $\widehat{n}_4 = 13.3$ and $\widehat{n}_5 = 1.1$ show that the lack of fit is primarily due to the underestimation of $n_5$, this cell contributes about 80% (14.6) to the total $X^2$ value. In terms of AIC the models $\mathcal{Z}_\beta$ and $\mathcal{Z}_{\gamma,\beta}$ fit best. Both models estimate the SP probability a little above 12%. The best model is $\mathcal{Z}_{\gamma,\beta}$, with the marginal distribution of the fitted values of $S_i$ given by

$$\widehat{\pi}(\mathcal{Z}_{\gamma,\beta}) = (0.657, 0.267, 0.063, 0.011, 0.002, 0.000).$$

TABLE 2
*Loglikelihoods, AIC's, BIC's and Pearson $X^2$ statistics, and SP estimates $\widehat{\theta}^*$*

|  | Model | Loglik. | AIC | BIC | $k$ | $X^2$ | $df$ | $\widehat{\theta}^*$ |
|---|---|---|---|---|---|---|---|---|
| $\mathcal{M}$ | Multinomial | $-1170.8$ | 2351.6 | 2375.4 | 5 | 6.1 | 0 | – |
| $\mathcal{P}$ | Poisson | $-1183.3$ | 2368.6 | 2373.4 | 1 | 56.0 | 5 | – |
| $\mathcal{Z}_0$ | ZIP (null) | $-1173.2$ | 2350.5 | 2360.0 | 2 | 19.6 | 4 | 0.126 |
| $\mathcal{Z}_\beta$ | ZIP (incl. $\beta$) | $-1167.0$ | 2348.1 | 2381.5 | 7 | – | – | 0.124 |
| $\mathcal{Z}_{\gamma,\beta}$ | ZIP (full) | $-1165.0$ | 2348.0 | 2391.0 | 9 | – | – | 0.121 |

The AIC and BIC disagree with respect to model choice. Since randomized response requires much larger samples than direct question designs and the BIC punishes for sample size, we feel that the BIC might be too conservative. Therefore, we prefer the model with the lowest AIC, which is $\mathcal{Z}_{\gamma,\beta}$. This choice is further motivated by the fact that the AIC decreases to 2343.4 when the four nonsignificant regression parameters in this model (see Table 3) are set to zero. In this case the BIC becomes 2367.2, so that according to this criterion, $\mathcal{Z}_0$ remains the preferred model. The disagreement between the two criteria indicates that the evidence for the effects of the predictors in model $\mathcal{Z}_{\gamma,\beta}$ is not strong.

Table 3 presents the parameter estimates of the predictors in model $\mathcal{Z}_{\gamma,\beta}$. The upper part of the table shows the results for the predictors in the vector $\mathbf{x}$. The last column reports the effect size $\exp(\widehat{\beta})$, expressing the relative change in the Poisson parameter for a unit change in the predictor. (For continuous variables, the standardized effect size can be computed by raising the reported effect size to the power of the standard deviation of the predictor.) The variables *year unemployed* and *knowledge rules* are significant predictors of the Poisson parameter. Regulatory noncompliance increases after the first year of unemployment; the estimated number of rule violations for a person unemployed longer than 1 year is 1.78 times that of a person unemployed less than 1 year. Better knowledge of the rules is associated with lower levels of regulatory noncompliance; the standardized effect size of 0.78 denotes the factor change in the Poisson parameter for each standard deviation increase in the score on *knowledge rules* (sd = 0.90).

The lower part of Table 3 reports the parameter estimates for the predictors in the vector $\mathbf{z}$. The last column reports the effect size $\exp(\widehat{\gamma})$, which expresses the relative change in the odds of SP for a unit change in the predictor. The parameter estimate for the variable *understanding* is significant. Better understanding of the forced response method results in less self-protective responses; the standardized effect indicates that the odds of SP decrease by approximately two-thirds (0.67) for each standard deviation increase in the score of *understanding* (sd = 0.85).

In order to assess the fit of model $\mathcal{Z}_{\gamma,\beta}$ more closely, we evaluate the correspondence between the observed and fitted frequencies within the response categories of each the predictor variables. Figure 3 plots the Pearson residuals $(n_{s^*x_{jk}} - \widehat{n}_{s^*x_{jk}})/\sqrt{\widehat{n}_{s^*x_{jk}}}$, with $n_{s^*x_{jk}}$ denoting the observed frequency of persons with sum score $s^*$ and score $k$ on predictor $j$, and $\widehat{n}_{s^*x_{jk}}$ denoting the corresponding fitted frequency. Because of the low frequencies of the observed sum scores 4 ($n = 7$) and 5 ($n = 5$), these two categories have been collapsed into the single sum score category 4/5.

TABLE 3
*Parameter estimates, standard errors (se), t-values and effect sizes for model $\mathcal{Z}_{\gamma,\beta}$*

| Predictors in x | $\widehat{\beta}$ | (se) | *t*-val. | $\exp(\widehat{\beta})$ |
|---|---|---|---|---|
| Constant | −0.13 | (0.38) | −0.32 | – |
| Gender (female) | 0.21 | (0.22) | 0.95 | 1.23 |
| Age ($< 26$) | 0.50 | (0.36) | 1.39 | 1.65 |
| Education | 0.19 | (0.18) | 1.07 | 1.21 |
| Year unemployment ($< 2004$) | 0.58 | (0.29) | 1.97 | 1.78 |
| Knowledge rules | −0.27 | (0.12) | −2.34 | 0.76 |
| **Predictors in z** | $\widehat{\gamma}$ | (se) | *t*-val. | $\exp(\widehat{\gamma})$ |
| Constant | −0.64 | (1.04) | −0.61 | – |
| Trust | 0.14 | (0.33) | 0.43 | 1.15 |
| Understanding | −0.46 | (0.23) | −1.99 | 0.63 |

The upper three plots of Figure 3 do not show systematic patterns or outliers for the dummy variables. Only the category of respondents younger than 27 with sum score zero shows a moderately large (negative) residual, indicating that this group is slightly overestimated. The remaining four plots in the lower part of Figure 3 show large residuals for the predictor *trust*. The plot shows underestimation of respondents who have no trust in the randomized response design and who have an observed sum score of either zero or four or five, with an exceptionally large residual in the latter category ($n = 3$). Since the combination of no trust in the randomized response design and *yes* responses to (almost) all sensitive questions is somewhat counterintuitive, it suggests the presence of a response mechanism opposite to that of SP; it may be the case that there are (a few) respondents who do not trust randomized response and who therefore (almost) always answer *yes*, irrespective of the outcome of dice. Although this is only a tentative explanation for the large residual, it would be interesting to see whether a similar response mechanism could also be detected in other randomized response applications.

The results presented in this section show evidence for the presence of SP in the data; the models with the SP parameter fit better than the other models. The degree of freedom needed to estimate the SP parameter is gained by the Poisson assumption for the sum scores of regulatory noncompliance. In this application the Poisson models fit the data well, except for the underestimation of the five cases with an observed sum score 5. Further research is needed to explore the nature of this misfit.

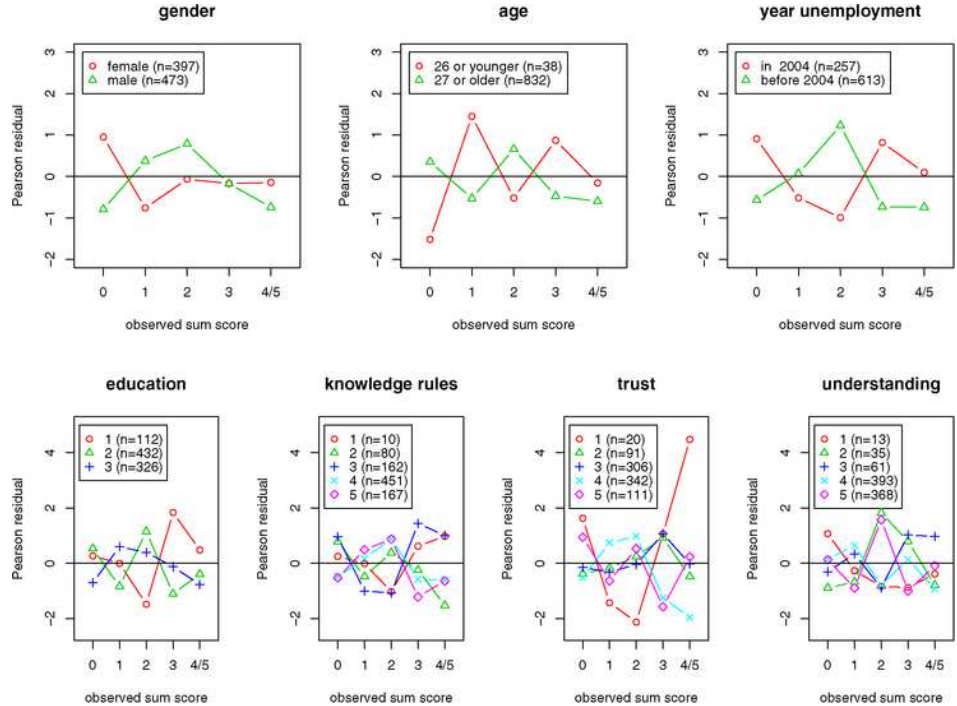

Fig. 3.   *Pearson residuals given predictor score and observed sum score (with sum scores 4 and 5 collapsed into single category).*

**5. Discussion.** In this paper we introduce a zero-inflated Poisson regression model for the analysis of randomized response sum score data. The central assumption underlying the model is that the true sum score variable follows a Poisson distribution, and that the presence of SP results in a zero-inflated distribution of the observed sum score variable. We present an example with a randomized response sum score variable assessing non-compliance with social security regulations. The ZIP randomized response model is used to find (1) the probability distribution of regulatory noncompliance, (2) the probability of SP, (3) significant predictors of regulatory noncompliance and (4) significant predictors of SP.

From a substantive point of view, the ZIP randomized response regression model yields some interesting results. For officials at the Department of Social Affaires, the negative effect of rule knowledge on regulatory noncompliance, suggesting that noncompliance is to a certain extent due to ignorance, may be of assistance in the formulation of new policies. The negative association between understanding of the forced response design and the probability of SP is especially interesting to social scientists interested in randomized response method. This result, that coincides with the conclusions of a study of psychological aspects of randomized response [Landsheer, van der Heijden and van Gils (1999)], suggests that adjustments in the instructions that would help the respondents to better understand the forced response design may reduce response bias and thereby enhance the validity of the responses.

The central assumption of the model that the true sum score variable is generated by a Poisson process is questionable since (1) the number of Bernoulli (i.e., binary randomized response) variables making up the randomized response sum score variable is limited and (2) the success probabilities (i.e., the prevalence of the sensitive characteristics) are not identical. Based on the univariate prevalence estimates, we demonstrate that in our example the Poisson approximation is satisfactory. An evaluation study with manipulation of the numbers of Bernoulli variables and of success probabilities (not reported here) shows that the quality of Poisson approximation is most affected if one (or more) of the success probabilities becomes larger. This finding is corroborated by the more formal result of Serfling (1978) that the absolute deviations between a series of Bernoulli variables with different success probabilities and its Poisson approximation increase as a function of the squared success probabilities. Although it is difficult to give exact figures, we feel that the Poisson assumption is justified as long as the prevalence estimates of the binary randomized response variables do not exceed 0.25. Since randomized response deals with sensitive characteristics that are rare by definition, the risk of this happening should be small.

## SUPPLEMENTARY MATERIAL

**The social security survey data** (doi: 10.1214/07-AOAS135SUPP; .pdf). The survey was conducted in 2004 by the Dutch Department of Social Affaires amongst 870 social security beneficiaries. The data contain the responses to five randomized response items assessing noncompliance with social security regulations, and seven background variables.

M. J. L. F. Cruyff
P. G. M. van der Heijden
Department of Methodology
  and Statistics
Utrecht University
3508 Utrecht
The Netherlands
E-mail: m.cruyff@uu.nl
       p.g.m.vanderheijden@uu.nl

U. Böckenholt
Department of Marketing
McGill University
Montreal
Canada
E-mail: ulf.bockenholt@mcgill.ca

A. van den Hout
MRC Biostatistics Unit
Cambridge
United Kingdom
E-mail: ardo.vandenhout@mrc-bsu.cam.ac.uk